\documentclass[12pt]{iopart}

\expandafter\let\csname equation*\endcsname\relax
\expandafter\let\csname endequation*\endcsname\relax
\usepackage{amsmath}
\usepackage[utf8]{inputenc}
\usepackage{graphics}		
\usepackage{graphicx}	
\usepackage{float} 
\usepackage{indentfirst}
\usepackage{array}
\usepackage{multirow}
\usepackage{xcolor}
\usepackage{soul}
\usepackage{cite}
\usepackage{mathtools}
\begin{document}

\title[MICROSCOPE T-SAGE characterization]{MICROSCOPE instrument in-flight characterization}

\author{Ratana Chhun$^1$, Emilie Hardy$^1$, Manuel Rodrigues$^1$, Pierre Touboul$^1$, Gilles M\'etris$^2$, Joel Berg\'e$^1$, Damien Boulanger$^1$, Bruno Christophe$^1$, Pascale Danto$^3$, Bernard Foulon$^1$, Pierre-Yves Guidotti$^3$ \footnote{Current address: AIRBUS Defence and Space, F-31402 Toulouse, France}, Phuong-Anh Huynh$^1$, Vincent Lebat$^1$, Fran\c{c}oise Liorzou$^1$, Alain Robert$^3$}

\address{$^1$ DPHY, ONERA, Université Paris Saclay, F-92322 Châtillon, France}
\address{$^2$ Universit\'e C\^ote d{'}Azur, Observatoire de la C\^ote d'Azur, CNRS, IRD, G\'eoazur, 250 avenue Albert Einstein, F-06560 Valbonne, France}
\address{$^3$ CNES, 18 avenue E Belin, F-31401 Toulouse, France}

\ead{ratana.chhun@onera.fr, gilles.metris@oca.eu, manuel.rodrigues@onera.fr}
\vspace{10pt}
\begin{indented}
\item[]January 2021
\end{indented}

\begin{abstract}
Since the MICROSCOPE instrument aims to measure accelerations as low as a few 10$^{-15}$\,m\,s$^{-2}$ and cannot operate on ground, it was necessary to have a large time dedicated to its characterization in flight. After its release and first operation, the characterization experiments covered all the aspects of the instrument design in order to consolidate the scientific measurements and the subsequent conclusions drawn from them. Over the course of the mission we validated the servo-control and even updated the PID control laws for each inertial sensor. Thanks to several dedicated experiments and the analysis of the instrument sensitivities, we have been able to identify a number of instrument characteristics such as biases, gold wire and electrostatic stiffnesses, non linearities, couplings and free motion ranges of the test-masses, which may first impact the scientific objective and secondly the analysis of the instrument good operation.
\end{abstract}

%
\vspace{2pc}
\noindent{\it Keywords}: MICROSCOPE, accelerometer, control laws, electrostatic stiffness, in-orbit characterization
%
%
%
%

\section{Introduction}

The goal of the Microscope mission is to test the \hl{Weak} Equivalence Principle in orbit, measuring and comparing the fall in the Earth gravitational field of two concentric masses \cite{rodriguescqg1} with an accuracy of $10^{-15}$. \hl{The payload called Twin-Space Accelerometer for Gravitation Experiment, T-SAGE, is made of two independent differential accelerometers, dedicated to the WEP.} \hl{Instead of letting one mass drift away from the other,} the accelerations needed to maintain \hl{the two masses of each differential accelerometer} on the same orbit \hl{are compared}. 
A non null differential signal collinear to Earth's gravity field would indicate a violation of the Equivalence Principle. \hl{The two differential accelerometers composing T-SAGE are called SU which stands for Sensor Unit. Each SU is composed of two accelerometers. Each accelerometer, also called Inertial Sensors (IS), is based on the principle of electrostatic levitation of a test-mass inside a cage of electrodes whose functions are to simultaneously detect the displacements of the test-mass and apply the electrostatic forces needed to maintain the test-mass at the center of the cage. In the case of MICROSCOPE, these test-masses are cylindrical to be able to maintain the two centers of gravity of a pair of test-masses on the same orbit by having one IS inside the other. Dimensions aside, the mechanics and electronics designs of all four IS are strictly identical.} One differential accelerometer, SUEP, is composed of two masses of different materials and provides measurements for the test itself. The other, SUREF, is composed of two masses of identical material to check the experiment since no violation signal is expected. \hl{In both SU, the test-mass of IS1, the internal IS, is made of a platinum alloy while the masses of the two IS2, the external IS, are made of different materials: the same platinum alloy for IS2-SUREF and a titanium alloy for IS2-SUEP. The complete apparatus, including the mechanics and the electronics, is much more detailed in} \cite{liorzoucqg2}. 
The test is performed several times over the course of the mission. Each test phase has its own calibration sub-phase, needed to check some parameters which could be used to correct the measurement data \cite{hardycqg6}. To consolidate these scientific measurements, several more technical sessions \cite{rodriguescqg4} have also been run to complement the knowledge of the T-SAGE instruments. Technical sessions are also dedicated to improve the control laws of the instruments during flight operation. The first observations of the instrument outputs were made during the commissioning phase at the start of the in-orbit operation. Before scientific measurements can be produced, the very first step consists in checking the instrument operability. \hl{Thus in section} \ref{section:release}, \hl{we will present the first moments of operation of the instrument from clamping to first switch on. In section} \ref{section:control}, \hl{we will see how the instrument control laws have been optimized in flight. Section} \ref{section:charac} \hl{describes the instrument characterization, starting with the stiffness which is an important characteristic in our electrostatic accelerometers. Next observations deal with the acceleration measurements main characteristics, namely bias and noise with a focus on their evolution. The quadratic factor estimation is then presented although it is not a major characteristic. We present the measurement of the free motion ranges of the masses, which brings some functional information on the instrument. Last but not least, we present the internal couplings of the instrument which play an important role in the final performance of the test.}


\section{Test-mass clamping, release and first control} \label{section:release}

\begin{figure}[h]
        \centering
        \includegraphics[scale=0.33]{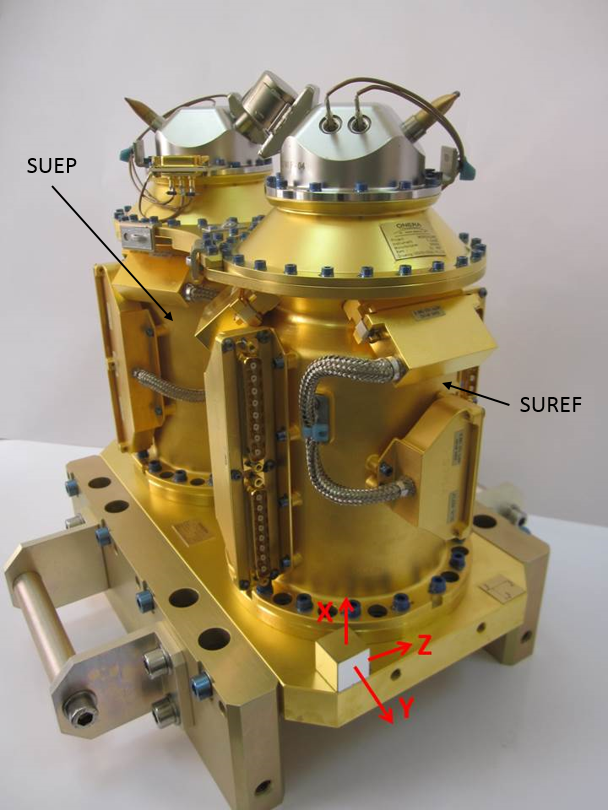}
        \includegraphics[scale=0.33]{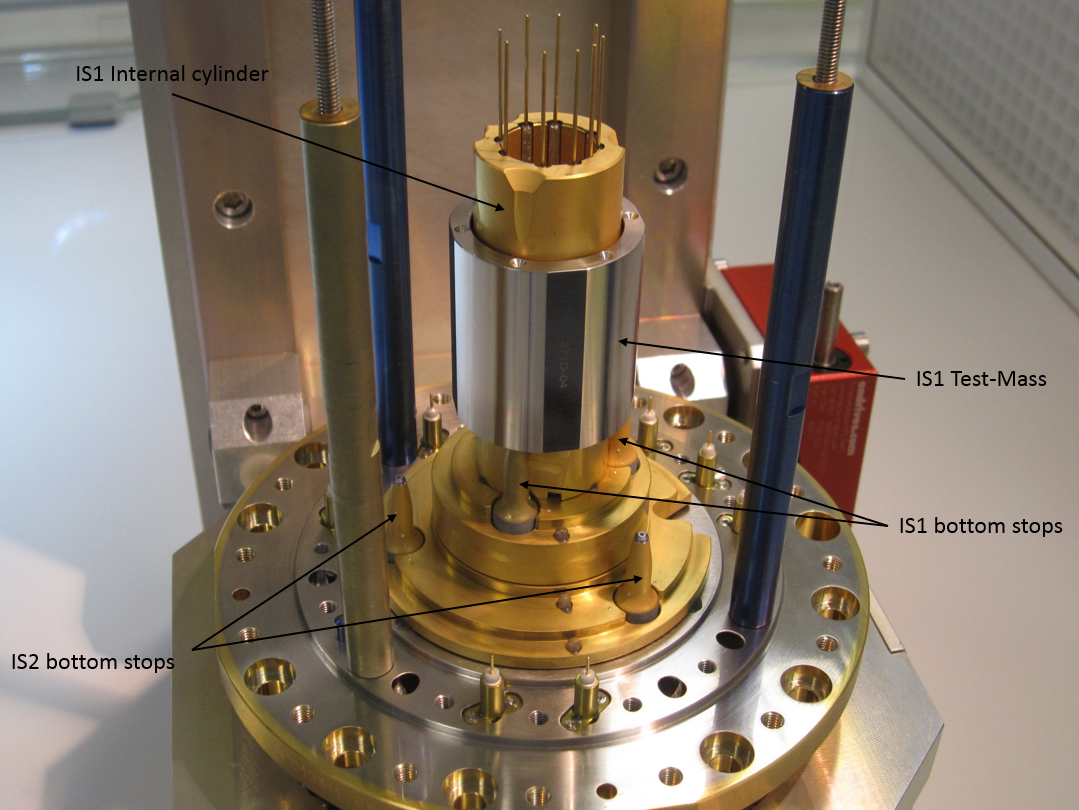}
        \includegraphics[scale=0.35]{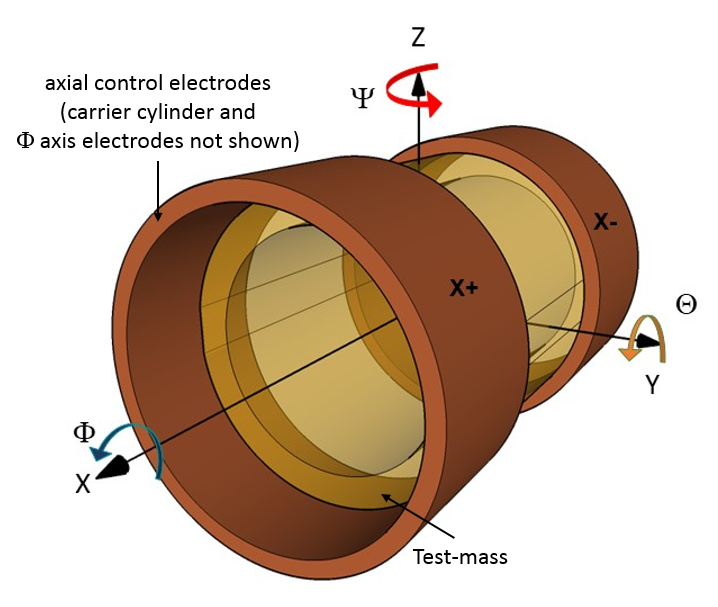}
        \includegraphics[scale=0.35]{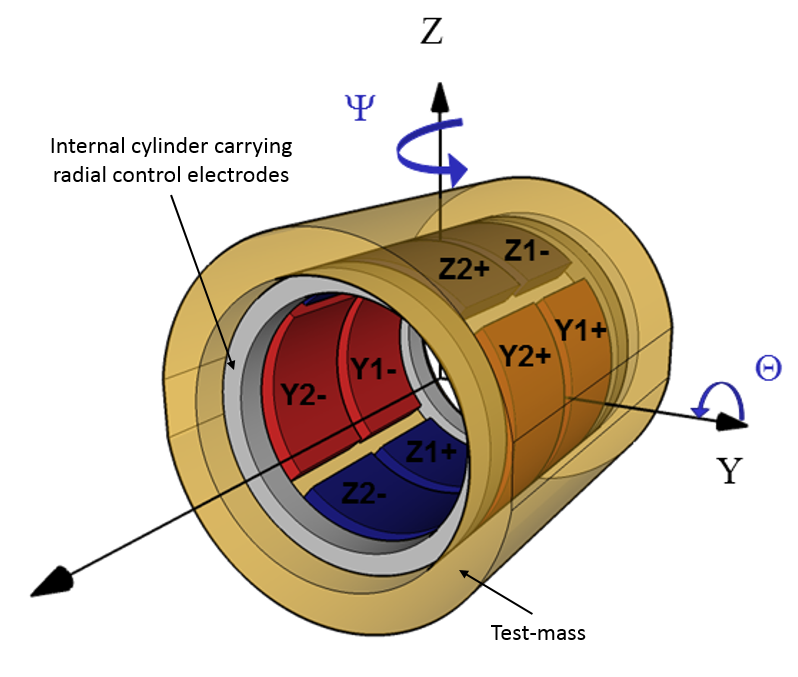}
        \caption{\hl{Top left: T-SAGE featuring its two SU, SUREF and SUEP. Top right: open core of one SU during integration showing inner test-mass resting on lower stops; upper clamp and outer IS not present. Bottom left: representation of one test-mass surrounded by the two $X$ axis control electrodes. Bottom right: representation of one test-mass surrounding the radial axes ($Y$, $Z$, $\Theta$ and $\Psi$) control electrodes.}}
        \label{fig:stops}
\end{figure}

Fearing damage at the core level of the instruments due to the launch vibrations, the whole instrument is maintained hyperstatic by clamping the four test-masses with a blocking mechanism consisting of 6 fingers or stops slightly inserted at each opposite test-mass faces (3 at each end as shown in Fig. \ref{fig:stops}). Once the satellite was in orbit on the 26$^{th}$ of April 2016 and well operating, the first switch-on of the four inertial sensors was realized on the 2$^{nd}$ of May 2016. The global commissioning phase started then.

\begin{figure}[h]
        \centering
        \includegraphics[scale=0.31]{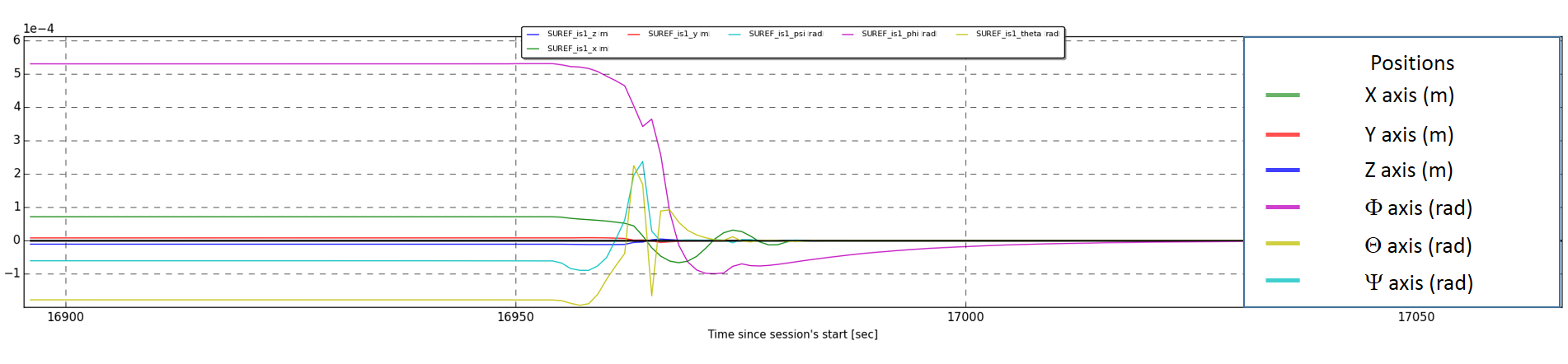}
        \includegraphics[scale=0.31]{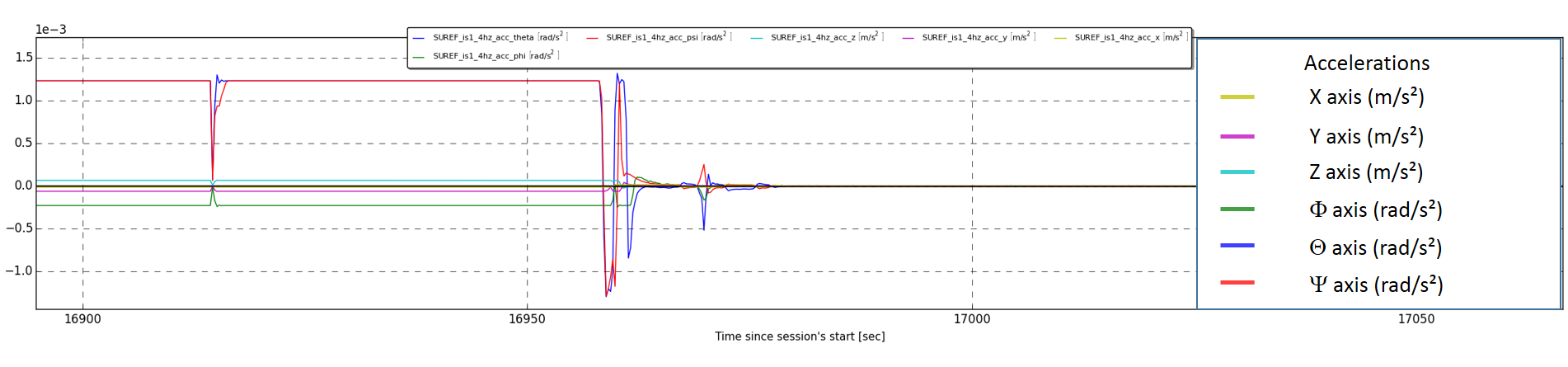}
        \caption{First control of IS1-SUREF (inner test-mass), accelerations and positions of all 6 degrees of freedom converge to zero at first switch on of the instrument}
        \label{fig:firstcontrol}
\end{figure}

After checking the thin 7$\mu$m gold wire connecting electrically the test-masses \cite{liorzoucqg2}, the very first operation on the instrument consisted in the simultaneous release of the four test-masses initially clamped during storage on Earth and launch, with the dread of one of the masses remaining stuck to one of the clamping fingers of the blocking system \hl{due to material molecular adhesion after several months before flight of storage in clamped configuration. This was mitigated during development by on-ground long duration tests on coatings to determine the electrostatic force necessary to separate clamped parts though.} The preliminary observation of the acceleration and position outputs in Fig. \ref{fig:firstcontrol} confirms the expected situation that the test-masses are blocked by their clamping fingers and pushed toward the top end of their free motion range along the axial $X$ axis, the control accelerations trying, in vain, to bring the masses back to their equilibrium position. On other axes, acceleration and position outputs are different from zero when taking into account machining and integration \hl{defects} for each test-mass. The general behavior is that it is pretty much aligned with its surrounding electrode cage, as deduced from observing the radial $Y$, $Z$, $\Theta$ and $\Psi$ axes, but twisted around its axial axis as observed from $\Phi$ axis measurement (the reference frame is described in Fig. \ref{fig:stops}). This is likely due to the misalignment between all 6 clamping fingers no matter which end they are at.

Once retracted by a few tens of microns, the fingers act as stops which limit the motion range of each test-mass. Indeed the fingers do not clamp anymore the test-mass but do remain slightly inserted to limit the mass motion along and around all axes simultaneously. The observation of their 48 acceleration and position outputs lifts all doubts: the test-masses are servo-controlled to the center of their respective electrode cages, as shown in Fig \ref{fig:firstcontrol} for the SUREF (IS1 for inner test-mass, IS2 for the outer test-mass). SUEP presents the same behavior. $\Phi$ rotation is the axis which converges the slowest. The $X$ axis displacement converges from its initial 75$\mu$m position. The radial axes $Y$, $Z$, $\Theta$ and $\Psi$ converge following a common pattern from an uncontrolled initial position. All these observations of the proof-mass meet the expected behavior of the servo-control performed by simulation.

Still subsequent frequency analyses of the acceleration outputs over the course of the following few months, have shown that the control laws initially defined before launch could be improved on several aspects.

\section{Control laws optimization} \label{section:control}

\begin{figure}[h]
        \centering
        \includegraphics[scale=0.45]{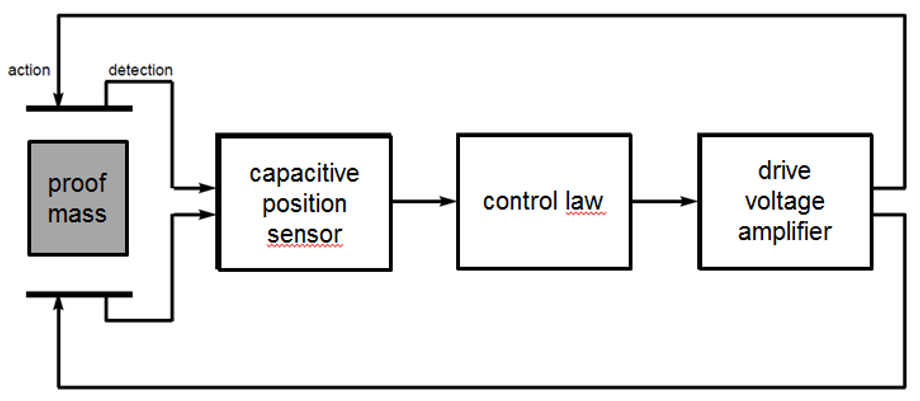}
        \caption{Representation of \hl{one test-mass degree of freedom} servo-control}
        \label{fig:loop}
\end{figure}

The electrostatic sensor control loop is schematized in Fig. \ref{fig:loop}. \hl{This is valid for any degree of freedom of any of the four Inertial Sensors.} The test-mass is surrounded by a set of electrodes, so that a test-mass motion induces a variation of capacitance between the mass and the electrodes in regard. This variation is measured by a capacitive detector whose output signal is digitized. By combining the outputs of the different electrode couples, the test-mass position and rotation are deduced along the six degrees of freedom. A corrector computes the voltage to be applied onto the electrodes in order to compensate for the test-mass motion and to keep it centered. The computed voltage is converted back into an analog signal before being amplified. An electrical potential is applied asymetrically by the electrodes in order to develop an actuation force opposed to the test-mass displacement. The electrostatic actuation keeps the test-mass motionless inside the satellite, and therefore provides a measurement of the acceleration necessary to compensate all other ``natural'' accelerations of the test-mass relatively to the satellite. Ref. \cite{liorzoucqg2} gives the detailed operation and equations of the electrostatic accelerometer along all axes.

\begin{figure}[h]
        \centering
        \includegraphics[scale=0.45]{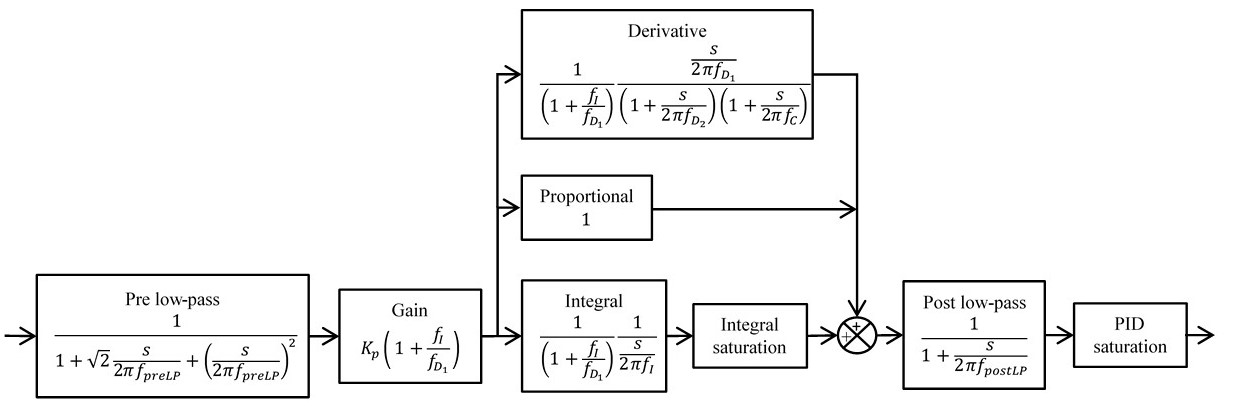}
        \caption{Representation of the controller, as implemented in the ICU. The value of the gain and frequencies are presented in table \ref{table:PID}}
        \label{fig:PID}
\end{figure}

The corrector is a PID-type (proportional-integral-derivative) controller implemented into a Digital Signal Processor (DSP). The digital controller operates at 1027\,Hz and delivers data to the satellite at a 4\,Hz rate.  The PID blocks are modelized by a parallel scheme as in Fig. \ref{fig:PID}. It is surrounded by a second order low-pass filter upstream and a first order low-pass filter downstream.

The choice of the control loop parameters is constrained by the electrostatic configuration, the mission operational specifications and the instrument performance objectives.

The satellite Drag-Free Attitude Control System (DFACS, \cite{rodriguescqg1, robertcqg3}) compensates the environmental perturbations applied to the satellite. The resulting residual acceleration affects in common mode both concentric test-masses. 
The DFACS is started with the accelerometers in their Full Range Mode (FRM) and performs a fine control of the satellite attitude making use of the star sensors in order to reduce the angular motion disturbances. Then the accelerometers are switched to their High Resolution Mode (HRM) in order to allow the DFACS to control the 6 degrees of freedom. Ref. \cite{robertcqg3} details the DFACS operation and performances. As the output of the accelerometers are used by the DFACS to control the satellite, the accelerometer servo-loop is designed to allow sufficient gain margins and stability to the DFACS  loop. The accelerometer range design takes into account the maximum applied acceleration in both modes (FRM and HRM) within the measurement bandwidth and takes into account some margins for out-of-the-band transients due for example, to  micro-meteorites, satellite \hl{microvibrations induced by MLI crackles under sun or Earth albedo heating} \cite{bergecqg8} and cold gas propulsion potential defects. Micrometeorite impacts are possible within the orbital environment, the accelerometer must sustain sudden shocks without the mass hitting the stops: the PID has been designed to counteract a debris momentum transfer of about $1.4\times{}10^{-3}$\,N\,s equivalent to a satellite instantaneous velocity variation of $5\times{}10^{-6}$\,{m/s}.  

Within the control range, the measurement range corresponds to the domain in which the output data provided by the instrument is not saturated. The predicted instrument noise and bias constrain the test-mass measurement range, as the sum of the bias and the noise spectral density integration shall fit with margin within the range.

The PID parameters for the control loop of the two identical internal test-masses of SUEP and SUREF have been defined while taking into account these requirements. The parameters of the external test-masses were then deduced by adjusting the control loop gain in proportion to the mass in order to keep the same bandwidth properties.

The parameter set has been tested on ground during the free-fall test of the instrument in the drop tower of the ZARM institute \cite{liorzoucqg2}, with a modified electronics configuration with respect to the flight configuration in order to take into account the short free-fall duration of less than 10 seconds and higher disturbing environment. Because of this modification leading to a higher range by a factor 4 than the flight FRM, the performances did not correspond to the in-flight performances. However it still allowed us testing the control loop dynamics and check the servo-controlling of the test-masses at the center of their electrostatic cages and assessing the validity of the chosen PID parameters.

\begin{figure}[h]
        \centering
        \includegraphics[scale=0.31]{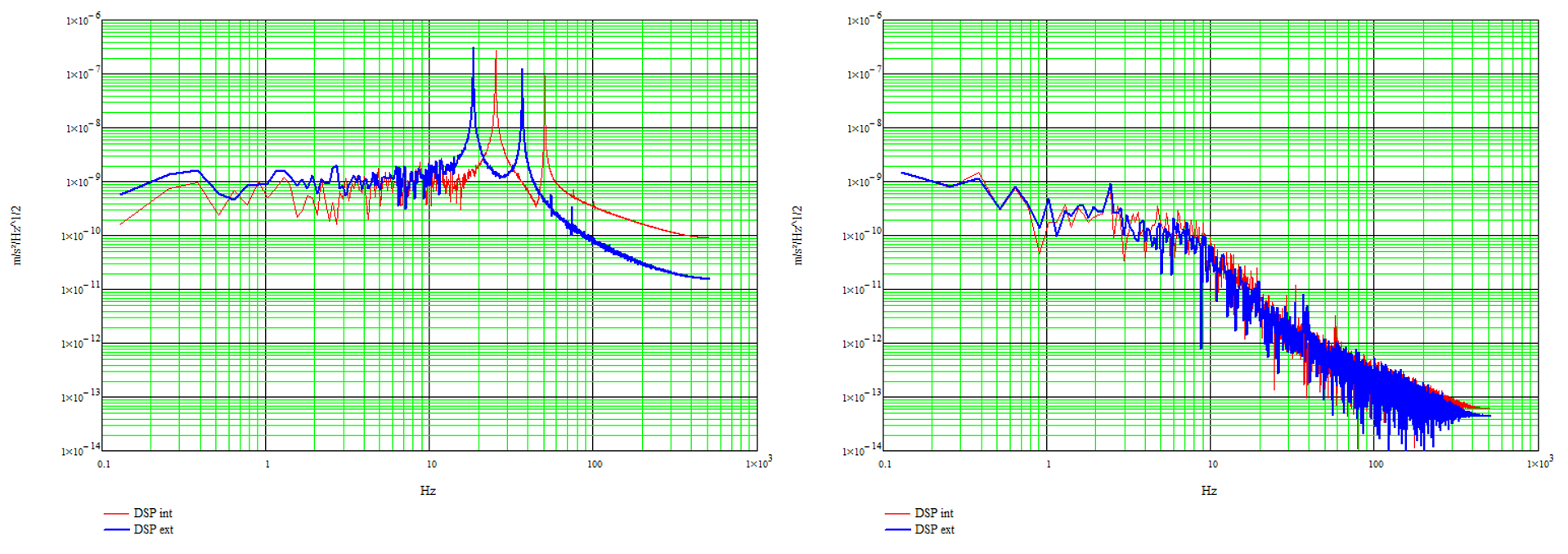}
        \caption{Acceleration spectrum density before (left)/after (right) the change of PID on SUEP $X$ axis, inner (red line) and outer (blue line) test-masses}
        \label{fig:PIDchange}
\end{figure}

\hl{In spite of the on-ground validation by simulation of the impact of the instrument control loop dynamics on the DFACS loop}, soon after the launch, the processing of the 1kHz measurement data of the test-masses accelerations, provided over 8\,s, demonstrated the presence of large spectral lines in the frequency range $[10-25]Hz$, out of the measurement frequency band. These resonances in the control loop (Fig. \ref{fig:PIDchange}) have a larger impact on SUEP than SUREF. These lines can disturb the calibration tests during which a stimulus signal needs to be injected into the DFACS or instrument loop. \hl{On-ground tests demonstrated the existence of these lines with test-masses not servo-controlled. An action/detection coupling was assumed to be the origin of these lines but they were deemed manageable at the time. But once in flight with servo-controlled test-masses, similar lines were observed as well, but at lower frequencies and with higher amplitude.} It was therefore decided to adjust the controller parameters in flight in order to decrease the amplitude of those spectral lines. This was achieved by decreasing the PID gain in this frequency range that leads to reduce the frequency bandwidth. For that purpose the performances of the PID controller have been computed for different sets of parameters. The set providing the maximal attenuation of the gain around 10Hz, while keeping acceptable values for the gain and phase margins, was selected. The drawback of this new PID adjustment is the phase shift higher than the requirement. As the accelerometers measurements are used in the DFACS loop, it results in a delay in the DFACS loop. However, thanks to the remarkable performances of the DFACS \cite{robertcqg3}, this additional delay remains acceptable.

The new PID adjustment was tested in flight during dedicated sessions, enabling us to confirm that the test-masses were servo-controlled at the center of their electrostatic cages and that the attenuation of the components at high frequencies was highly improved (see Fig. \ref{fig:PIDchange}).

Each axis of the four test-masses has its own well-tuned PID. The final set of parameters is presented in Tab. \ref{table:PID}.

\begin{table}[h]
\begin{center}
\begin{tabular}{ m{8.2cm} c c c c }
\hline
 Axis & $X$ & $Y/Z$ & $\Phi$ & $\Theta/\Psi$ \\ 
 \hline\hline
 Upstream cutoff frequency $f_{PreLP}$ (Hz) & 10 & 30 & 10 & 15 \\  
 Gain $K_p$ for SUEP and SUREF internal test-masses & -2 & 18.5 & -35.2923 & 50.87 \\
 Gain $K_p$ for SUEP external test-mass & -1.9957 & 2.22 & -1.7222 & 5.86 \\
 Gain $K_p$ for SUREF external test-mass & -3.0025 & 5.0129 & -7.7635 & 13.1390 \\
 Integral cutoff frequency $f_I$ (Hz) & 0.004 & 0.1 & 0.01 & 0.1 \\
 Low derivative cutoff frequency $f_{D1}$ (Hz) & 0.05 & 0.25 & 0.08 & 0.25 \\
 High derivative cutoff frequency $f_{D2}$ (Hz) & 1 & 20 & 5 & 7 \\
 Low-pass cutoff frequency $f_C$ (Hz) & 5 & 25 & 5 & 10 \\
 Downstream cutoff frequency $f_{PostLP}$ (Hz) & 10 & 30 & 10 & 15 \\
 \hline
\end{tabular}
\caption{Parameters of the instrument control loop PID controller (see Fig. \ref{fig:PID}) for the four test-masses in HRM mode}
\label{table:PID}
\end{center}
\end{table}

\section{Instrument characterization} \label{section:charac}
		
The mission did not only consist of scientific measurements. In order to improve the knowledge on all the aspects of the instrument and potentially understand a number of observations on the measurements, several characteristics of the electrostatic accelerometers were investigated through dedicated sessions. These characteristics all appear in the following simplified measurement equation of the sensor:

\begin{equation} \label{eq:1}
\ \Gamma_{meas_i}=\Gamma_{0_i}+K_{1_i}\Gamma_{app_i}+\frac{k_i}{m}d_i+K_{2_i}\Gamma_{app_i}^2+\sum_{j} Ct_{j,i}\Gamma_{app_j}+\sum_{j} Cr_{j,i}\dot{\Omega}_{j}+n_i\
\end{equation}

$\Gamma_{0_i}$ is the measurement bias along ``$i$'' axis, \hl{which includes a DC component and very low-frequency variations such as drift, acceleration systematic errors due to various defects in the mechanics or electronics, parasitic effects, which may even vary at the test frequency}. $\Gamma_{app_i}$ is the actual acceleration close to $\Gamma_{meas_i}$ by a \hl{relative} scale factor $K_{1_i}$ \hl{, close to 1, which is the ratio between the actual scale factor, converting voltage to acceleration, and its on-ground estimation. The error is due to the knowledge on the scale factor electrostatic model itself and to the mechanics and electronics manufacture accuracy (\textless1\%)}. $k_i$ is the stiffness inducing an additional force when the test-mass is displaced at a distance $d_i$ with respect to its equilibrium position ($m$ being the mass of the test-mass). $K_{2_i}$ is the quadratic factor. $Ct_{j,i}$ represent the couplings with the linear accelerations $\Gamma_{app_j}$ along ``$j$'' axes. $Cr_{j,i}$ represent the couplings with the angular accelerations $\dot{\Omega}_{j}$ about ``$j$'' axes. \hl{The last term} $n_i$ represents the measurement noise.

	\subsection{Mass displacement and stiffness characterization} \label{section:stiffness}
	
One important characteristic of the electrostatic accelerometer is the electrostatic stiffness. This characteristic induces into the acceleration measurement a bias proportional to an offcentering of the test-mass with respect to its electrostatic equilibrium or zero position.

\begin{figure}[h]
        \centering
        \includegraphics[scale=0.39]{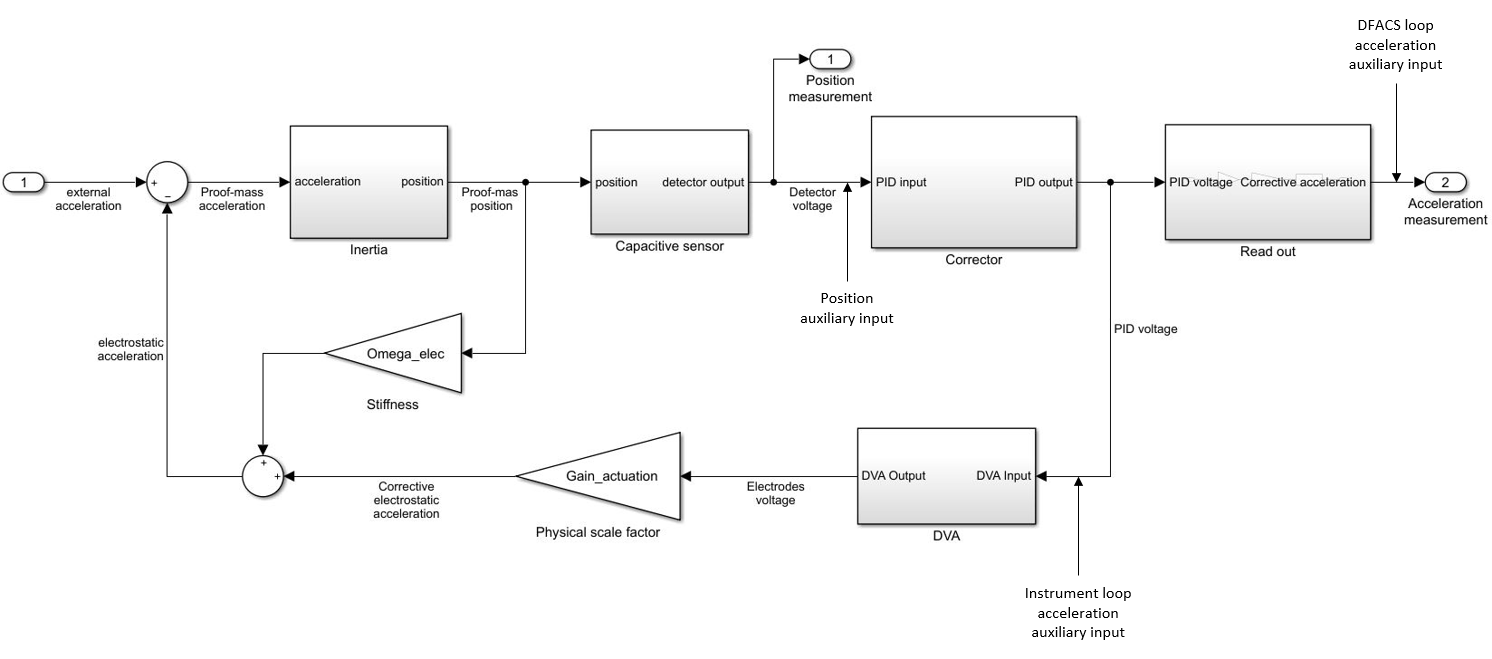}
        \caption{Representation of the instrument control loop in the Simulink tool}
        \label{fig:looplink}
\end{figure}

By introducing \hl{a position signal input} (Fig. \ref{fig:looplink})\hl{, sine-shaped around zero,} successively along each test-mass axis (Eq. (\ref{eq:displacement})), \hl{forcing the mass to move according to the input}, it is possible to measure a periodic acceleration proportional to the motion because of the stiffness.

\begin{equation} \label{eq:displacement}
d_{i}(t) = d_{i0} \sin(2\pi f_{\rm cal} t + \psi) , d_{\ne i}=0 
\end{equation}

\begin{figure}[h]
        \centering
        \includegraphics[scale=0.3]{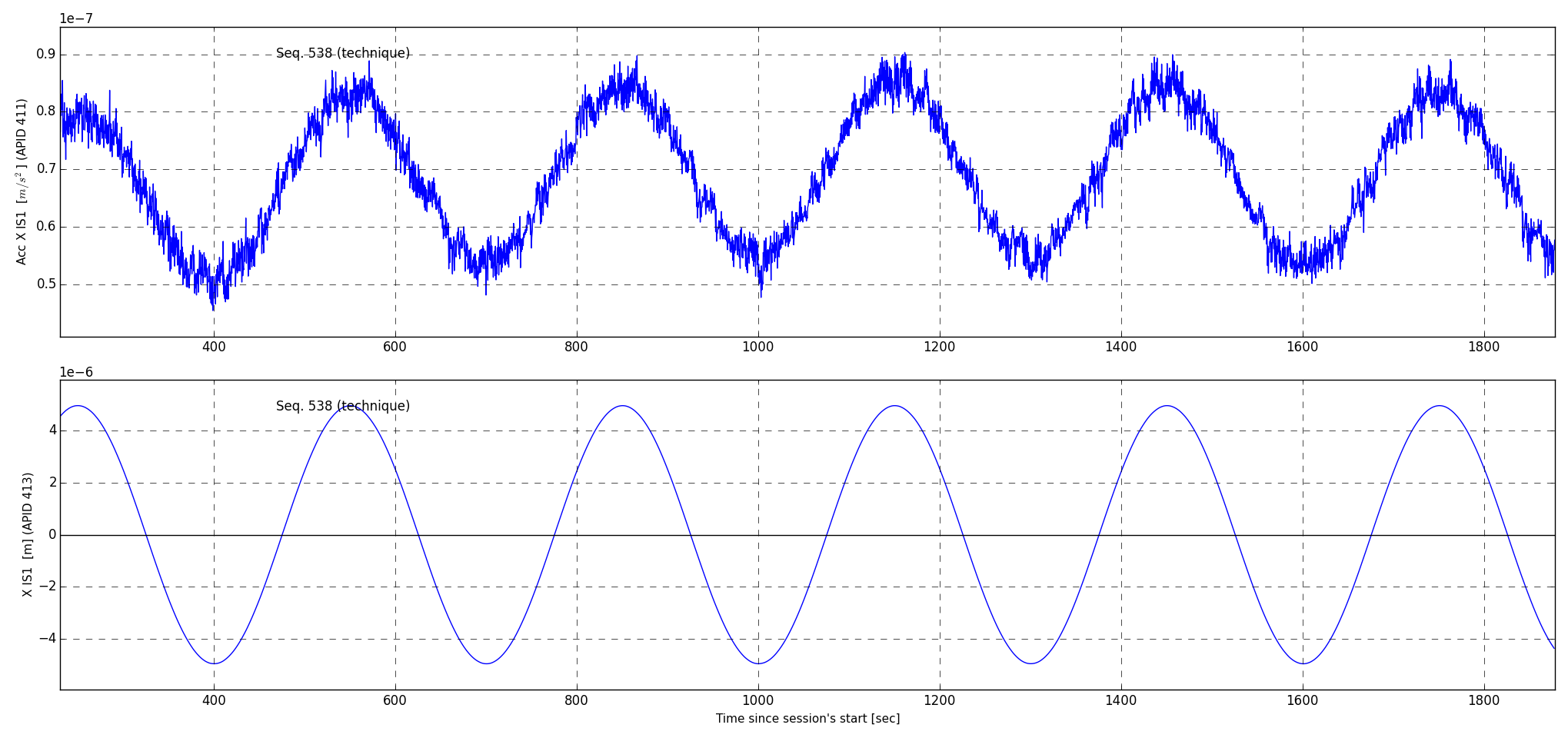}
        \caption{Stiffness measurement on SUEP IS1 $X$ (axial) axis in High Resolution Mode: a sine signal position input sets the mass in motion, displacement and corresponding acceleration are compared to obtain the stiffness, signals are in phase, interpreted as main gold wire influence}
        \label{fig:stiffX}
\end{figure}

\begin{figure}[h]
        \centering
        \includegraphics[scale=0.3]{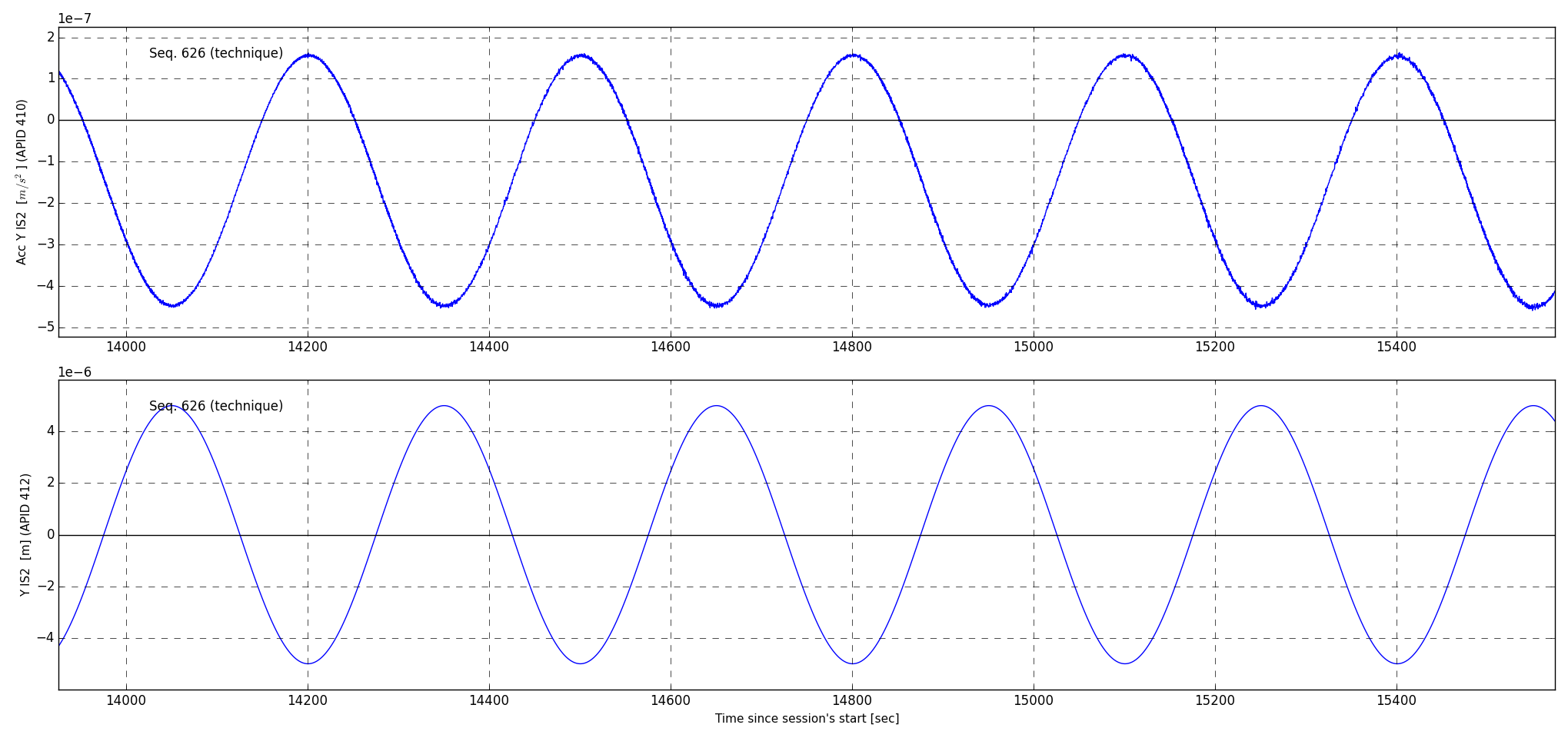}
        \caption{Stiffness measurement on SUREF IS2 $Y$ (radial) axis in High Resolution Mode: a sine signal position input sets the mass in motion, displacement and corresponding measured acceleration are compared to obtain the stiffness, signals are in opposition, interpreted as main electrostatic influence}
        \label{fig:stiffY}
\end{figure}

The stiffness was assessed in flight during the commissioning phase then assessed again at the end of the mission with a different electrical configuration (see Fig. \ref{fig:stiffX} and \ref{fig:stiffY}) using a least-square method on the displacement and the corresponding acceleration sines in Eq. (\ref{eq:displacement}) and  (\ref{eq:stiffmeas}).

\begin{equation} \label{eq:stiffmeas}
\Gamma_{i} = \Gamma_{exc_i} + \Gamma_{0_i} + \frac{k_i}{m} {d_i}
\end{equation}
where

\begin{equation} \label{eq:stiffmeas2}
\Gamma_{exc_i} = [T-In]_{i,i}{d_i} -\ddot{d_i} 
\end{equation}

$\Gamma_{exc_i}$ includes the kinematic effect of the displacement in the satellite reference frame and the Earth gravity gradient effect: the definition of the matrix of inertia [In] and of the gravity gradient matrix [T] is detailed in Ref. \cite{rodriguescqg1}.  At the oscillation frequency $f_{\rm cal}$, the angular velocity $\Omega$ is inferior to $4\times 10^{-5} rad/s$. Considering the mass excenterings (a few $\mu m$), the effect of [In] on the acceleration measurement is negligible with respect to the stiffness effect. The term in $\dot{d}_i$ is missing in Eq. (\ref{eq:stiffmeas2}) because the resulting effect is orthogonal to the ``$i$'' axis. Disturbing terms could potentially occur with the DFACS residual angular acceleration lower than $4\times 10^{-8} rad/s^2$ at $f_{\rm cal}$ and by considering the static offcentering of a few tens of micrometers, but this effect leads to a negligible effect.  The same can be said of the Earth gravity gradient. Only $\ddot{d_i}$ (about $2\times 10^{-9}m/s^{-2}$) has to be considered and is indeed corrected from the measurement before stiffness assessment. The results in Tab. \ref{table:stiffness} take these considerations into account.

By design the electrostatic stiffness, or torsion constant in rotation, is expected to be very low on the sensitive axes $X$ and $\Phi$ because of the principle of capacitive sensing on these axes based on the variation of \hl{surface overlapping electrodes and test-mass, as opposed to the variation of distance, or gap, between electrodes and test-mass, basis of the capactive sensing on radial axes, $Y$, $Z$, $\Theta$ and $\Psi$. Theoretically this overlap principle does not generate an electrostatic stiffness. But to be accurate, the control of $\Phi$ axis is a mix of gap and overlap variations} \cite{liorzoucqg2} \hl{, which, depending on the balance between electrostatic and gold wire stiffnesses, may generate a non null stiffness, either positive or negative.} However the actual stiffness measurements (see Tab. \ref{table:stiffness}) turn out being \hl{more significant} than expected (except around $\Phi$ axis of IS1-SUREF). Do note that these results, while approximated with respect to \cite{bergecqg10} remain comparable and compatible.

Along the $X$ axis this is likely due to the gold wire which induces a mechanical stiffness. The influence of the gold wire is indicated by the positive sign of the stiffness indicating a stabilizing effect in reaction to a displacement of the test-mass. On the external mass of SUREF, the effect of the gold wire is very strong while the torsion constants around $\Phi$ axis are negative on the masses of SUEP, which means the electrostatic stiffness is greater than expected, and compensates the effect of the gold wire. The internal mass of SUREF is the only one featuring a stiffness compatible with the theory. From one mass to another, the stiffness due to the gold wire is subject to vary a lot because it is dependant on its mechanical integration inside the core. Given the nature of the $7{\mu}m$ diameter gold wire and the procedure to glue it by both ends, the mechanical behavior of the gold wire when the mass is displaced is quite difficult to predict beforehand and to reproduce. By considering the free motion ranges presented in Tab. \ref{table:range}, especially in SUEP, it can also be argued that the proof-mass is quite close to the stops on one end. A stiffness due to the contact potential difference or patch effect may add itself up in the measurement, combined with this proximity if we pessimistically consider patch effects 3 times greater than initially predicted, $50mV$ instead of $15mV$. In SUREF, these patch effects would have to be 20 times greater, which is highly unlikely.

\begin{table}[h]
\begin{center}
\begin{tabular}{ m{3cm} c c c c }
\hline
Axis &IS1-SUREF &IS2-SUREF & IS1-SUEP & IS2-SUEP \\
 \hline\hline
 \multirow{2}{1cm}{X$({\times}10^{-3}$N\,m$^{-1})$} & 0.84$\pm$0.01 & 4.42$\pm$0.01 & 1.40$\pm$0.01 & 0.64$\pm$0.01 \\
& (0.01) & (0.01) & (0.01) & (0.01) \\
 \multirow{2}{1cm}{Y$({\times}10^{-2}$N\,m$^{-1})$} & -1.52$\pm$0.01 & -8.17$\pm$0.01 & -1.50$\pm$0.01 & -6.42$\pm$0.01 \\
 & (-2.02) & (-9.55) & (-2.02) & (-9.10) \\
 \multirow{2}{1cm}{Z$({\times}10^{-2}$N\,m$^{-1})$} & -1.52$\pm$0.01 & -7.15$\pm$0.01 & -1.48$\pm$0.01 & -6.31$\pm$0.01 \\
& (-2.02) & (-9.55) & (-2.02) & (-9.10) \\
\hline
 \multirow{2}{1cm}{$\Phi ({\times}10^{-8}$N\,m\,rd$^{-1})$} & 0.0$\pm$0.3 & 428.9$\pm$0.9 & -1.5$\pm$0.9 & -10.0$\pm$0.5 \\
& (-0.25) & (-0.77) & (-0.25) & (-0.77) \\
 \multirow{2}{1cm}{$\Theta ({\times}10^{-6}$N\,m\,rd$^{-1})$} & -1.75$\pm$0.01 & -34.37$\pm$0.03 & -1.55$\pm$0.01 & -30.06$\pm$0.01 \\
& (-2.13) & (-42.8) & (-2.13) & (-41.2) \\
 \multirow{2}{1cm}{$\Psi ({\times}10^{-6}$N\,m\,rd$^{-1})$} & -1.85$\pm$0.01 & -40.85$\pm$0.01 & -1.82$\pm$0.01 & -30.08$\pm$0.01 \\
& (-2.13) & (-42.8) & (-2.13) & (-41.2) \\
 \hline
\end{tabular}
\caption{Measured and expected (between brackets) stiffnesses and torsion constants in High Resolution Mode; theoretical values have been computed before flight assuming a perfect and simple electrostatic configuration. \hl{Considering the control principle, theoretical value along X is expected to be null but is actually adjusted by a cap value issued from an analytical model based on geometry and defects}}
\label{table:stiffness}
\end{center}
\end{table}

On the radial axes $Y$, $Z$, $\Theta$ and $\Psi$, the capacitive sensing is based on the variation of distance between test-mass and electrode thus generating a negative stiffness which tends to destabilize the equilibrium of the test-mass as opposed to the gold wire stiffness effect. On this basis, the stiffness only depends on the geometrical configuration and the voltages applied on the electrodes and the test-mass. The results obtained (see Tab. \ref{table:stiffness}) are quite comparable between the two internal masses of each SU since their electrostatic configurations are identical. They are slightly different between the two external masses because the voltages applied on the electrodes are also slightly different. The discrepancy between the results and the expected values is explained by the simplification of the theoretical electrostatic model and configuration.

The electrostatic main origin of the stiffness on the radial axes is also pointed out by the differences between the values measured during the campaigns led in High Resolution Mode and in Full Range Mode, meaning the polarization voltage applied on the mass is different with respect to the mode (respectively 5V and 40V). In Full Range Mode, the stiffnesses see their amplitude increased although not by a factor strictly equal to the square of the potential difference between the mass and the electrodes as the gold wire also has an influence independent of the electronics configuration albeit less impacting.

On the contrary, the stiffness along $X$ axis and the torsion constant around $\Phi$ axis are hardly affected by the change of the polarization voltage value, indicating the electrostatic part represents a minor contribution to the global stiffness essentially due to the gold wire.

\subsection{Measurement bias}

\begin{figure}[h]
        \centering
        \includegraphics[scale=0.46]{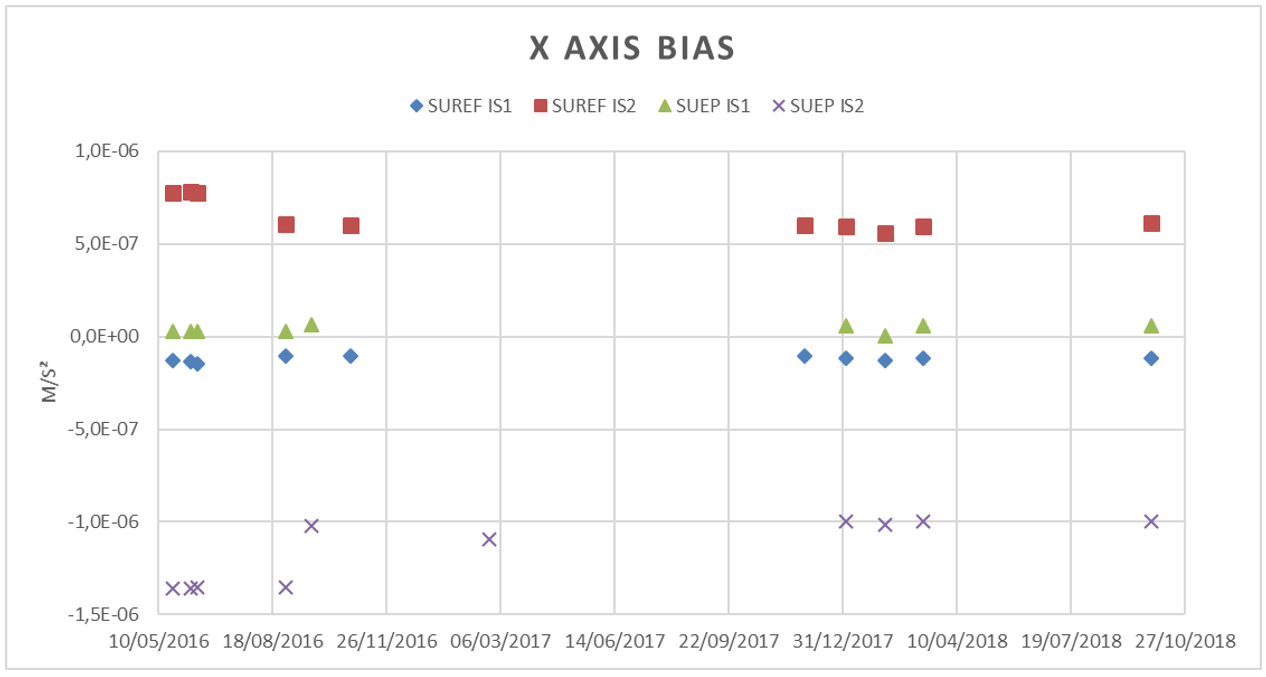}
        \caption{Bias evolution over the course of the 2-year-long mission along the $X$ axis of each inertial sensor}
        \label{fig:bias}
\end{figure}

The bias of each axis is periodically measured in order to be fed to the DFACS and minimize gas consumption among purposes, by biasing the DFACS so that it does not pointlessly compensate the instrument linear biases. Fig. \ref{fig:bias} presents the evolution of the bias along the $X$ axis of each inertial sensor. For this axis but more globally along all axes, the bias is quite stable over the whole mission, especially over the second half of the mission when the electronics configuration has attained its pretty much final form as opposed to the first half and especially around the first few months of the mission, when the biases were prone to variation as we were still adjusting the electronics. This is particularly observable towards the end of August 2016 on the external SUREF test-masses.

\subsection{Noise evolution}

\begin{figure}[h]
        \centering
        \includegraphics[scale=0.44]{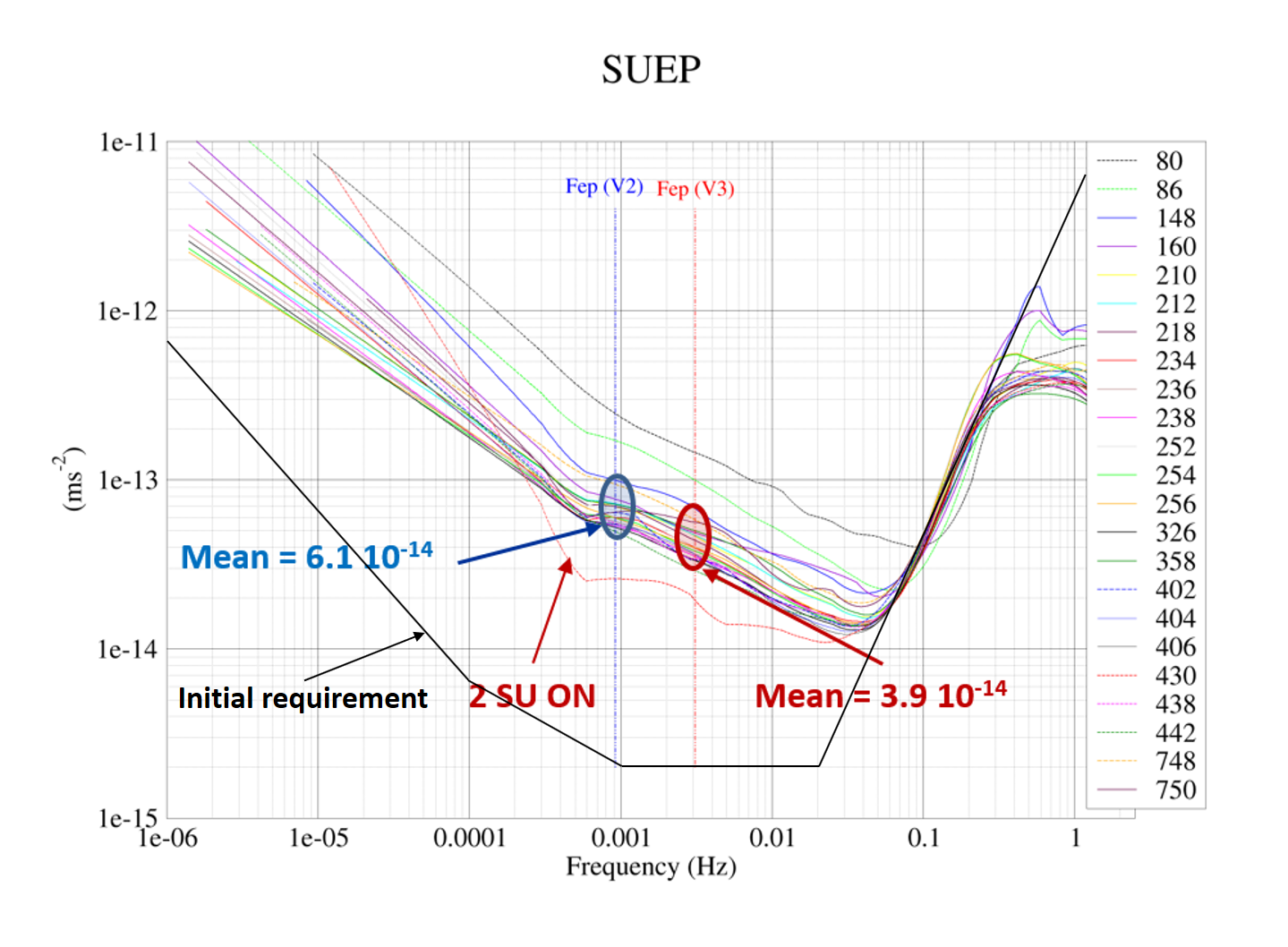}
        \caption{\hl{Smoothed} \cite{bergecqg7} \hl{amplitude of the FFT of the noise of} all EP test sessions with SUEP, normalized to 120 orbits \hl{by applying a factor $\sqrt{\text{duration in orbits}/120}$}}
        \label{fig:noisefft}
\end{figure}

The test of the equivalence principle has been performed several times over the course of the mission. The test performance is limited for an important part by the noise of the acceleration measurement. The spectra of the residual accelerations from all the test sessions, with the gravity field signal removed, have been plotted to compare the quality of the different sessions. Fig.  \ref{fig:noisefft} shows the spectra for the SUEP, smoothed using sliding frequency windows and linear interpolation between the central points of linear regressions applied on each of these windows. Since the duration of the sessions varies a lot from one to another, they are all normalized to 120 orbits \hl{by applying a factor $\sqrt{\text{duration in orbits}/120}$}. \hl{The origin of the noise has been discussed in} \cite{touboul19}. \hl{The low frequency range corresponds to the drift of acceleration due to the instrument sensitivities to environment components such as temperature. The high frequency range is as expected. The medium part of the bandwidth is higher than expected and considered due to the damping of the gold wire mainly. This paper focuses on the noise stability over the course of the mission though.}

In the measurement bandwidth $[10^{-4}-10^{-2}]Hz$, we can observe that all but 3 sessions have a similar spectrum level in Fig. \ref{fig:noisefft}. The first 2 sessions, numbers 80 and 86, feature a higher noise level than the pack, the very first test being the noisier. Session 430 is characterized by a different configuration as both differential accelerometers are switched on and thus the operating temperature is higher. Albeit shorter in duration, once normalized, the spectrum is much lower than in the other sessions.

By comparing the noise value at a given frequency for every session, it is even more apparent that, over the course of 2 years of mission, the noise level decreases while the spectrum keeps the same shape.

The improvement starts at the time the measurement frequency is chosen to be higher in rotating mode, around $3\times10^{-3}Hz$ instead of around $10^{-3}Hz$. But it is not established whether the improvement is due to this new configuration or to environment and more general mission conditions less favorable during the first months of the scientific mission.

This behavior is observable on both SUREF and SUEP accelerometers. The improvement starts around the same time, thus rejecting the potential influence of the change of the control laws, which occurred at very different dates for each SU. Improvement of noise with time has been observed in Lisa Pathfinder \cite{armano18} due to pressure diminution with time. In MICROSCOPE, the core is enclosed  in a hermetic vessel and cannot take advantage of a reduction of pressure due to the residual gas escaping to space vacuum as in Lisa Pathfinder. It is probably due to a different phenomenon.

	\subsection{Quadratic factor}

When dissymmetries of geometrical or electronic nature are taken into account, a disturbance acceleration depending on the square of the acceleration applied on the electrodes has to be considered. This term is proportional to the coefficient defined as the quadratic factor, $K_{2}$.

\begin{figure}[h]
        \centering
        \includegraphics[scale=0.75]{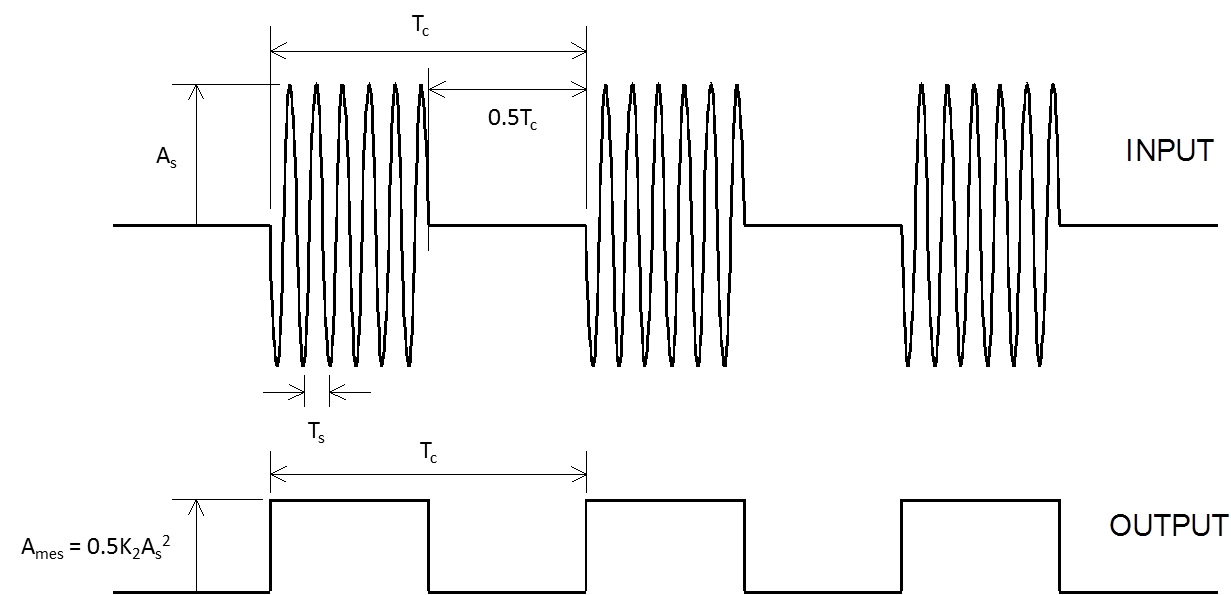}
        \caption{$K_2$ measurement: expected acceleration input and output signals}
        \label{fig:K2}
\end{figure}

\begin{figure}[h]
        \centering
        \includegraphics[scale=0.75]{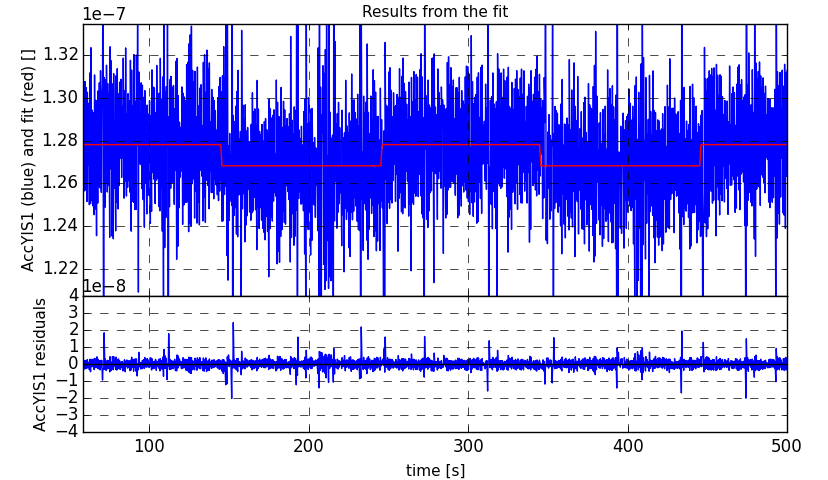}
        \caption{$K_2$ measurement on IS1-SUREF $Y$ (radial) axis: a square modulated sine signal input in acceleration generates a square signal, the amplitude is proportional to $K_2$, identified by a least-square fit}
        \label{fig:K2fit}
\end{figure}

To estimate this term, a periodic acceleration signal is input inside the accelerometer control loop (along the investigated degree of freedom, see Fig. \ref{fig:looplink}) in order to generate a periodic term due to $K_2$  (see Fig. \ref{fig:K2}). This signal is based upon a sine signal at a frequency $f_s$ higher than the measurement bandwidth and of amplitude $A_s$, modulated by a square signal of which the frequency $f_c$ is well inside the bandwidth. Due to $K_2$, a signal proportional to it is generated and its components are at $2f_s$ and $DC$. By selecting $f_s$ outside the measurement bandwidth, both components at $f_s$ and $2f_s$ are not visible in the output measurement. But by modulating the sine signal with the square signal, the $DC$ component is measurable in the output as a periodic square signal at frequency $f_c$ and of amplitude proportional to $K_2$ and the square of $A_s$  as expressed in Eq. (\ref{eq:K2}).  Fig. \ref{fig:K2fit} shows a measurement example and the resulting fit by a square signal estimated by a least-square method. The summary results for the linear axes are presented in Tab. \ref{table:K2}.

\begin{equation} \label{eq:K2}
\ \Gamma_{meas}=\Gamma_0+\frac{1}{2}K_2A_{s}^2\
\end{equation}

\begin{table}[h]
\begin{center}
\begin{tabular}{ m{2cm} c c c c }
\hline
Axis &IS1-SUREF &IS2-SUREF & IS1-SUEP & IS2-SUEP \\
 \hline\hline
 \multirow{2}{1cm}{$X (s^2/m)$} & $2520\pm796$ & $1250\pm868$ & $7260\pm344$ & $565\pm317$ \\
& (8000) & (20000) & (8000) & (20000) \\
 \multirow{2}{1cm}{$Y (s^2/m)$} & 770$\pm$27 & -120$\pm$4 & N/A & N/A \\
 & (6500) & (6500) & (6500) & (6500) \\
 \multirow{2}{1cm}{$Z (s^2/m)$} & -5560$\pm$313 & 22.5$\pm$9.4 & N/A & N/A \\
& (6500) & (6500) & (6500) & (6500) \\
 \hline
\end{tabular}
\caption{Quadratic factors estimated with a least-square method compared to specification between brackets, SUEP radial axes measurements did not give any significant result}
\label{table:K2}
\end{center}
\end{table}

	\subsection{Motion ranges} \label{section:ranges}

		For calibration purpose it is necessary to displace one of the test-mass along or around one of its degree of freedom by a few microns. Dedicated sessions have been set up in order to estimate the range margins for the amplitude of the displacement and to better evaluate the disturbances related to the distance test-mass/finger stops. \hl{The results are presented in Tab.} \ref{table:range}.
		
		The clamping fingers used during the launch are retracted from their initial position at +75$\mu$m for SUEP and +90$\mu$m for SUREF to a new theroretical position of -75$\mu$m. Thus after launch these stops limit the motion range of the mass to prevent any electrical contact with the electrodes \cite{liorzoucqg2}.  Because of the ball-point shape of the finger stops tips and the cone-shaped holes on the test-mass section facing the stops, and because of the accuracy of the realization and the assembly of these parts, the range can be \hl{modified by several tens of microns with respect to the expectation. In detail, part of the reduction is due to the integration process. The assembly of the six stops of each test-mass has to garantee their alignment with respect to their corresponding holes with their own uncertainties. Each stop is 4cm long and expectedly only 75$\mu$m away from the test-mass. A budget of free-motion range was performed before integration but could not be directly verified once the SU core got closed shut. In spite of accurate manufacture of each individual mechanical part, some reduction of the free motion was expected as presented in the budget in Tab.} \ref{table:range}. \hl{The capacity of the test-mass to move freely was nevertheless indirectly verified during free-fall tests} \cite{liorzoucqg2}.

On top of that the capacitive equilibrium due to the servo-control may be different from the geometrical equilibrium. \hl{Indeed all test-masses and electrode cylinders feature chamfers. The dissymetry of realization of these quickly offsets the test-mass in order to maintain capacitive equilibrium and thus reduces the distance between the test-mass and one set of stops (top or bottom).

Finally, to prevent the blocking mechanism from getting stuck during retractation, a certain freedom of movement is given to the plate supporting the bottom stops (the ones which retract), potentially adding misalignement of the stops with respect to the test-mass and further reducing the free motion range.}

The dedicated sessions for motion range evaluation have been performed in orbit with a triangle signal injected as a position secondary input (Fig. \ref{fig:looplink}) along or around each axis successively for each mass. The input signal starts from zero and takes 5 minutes to reach a peak which is set to a value over the maximum of the detector range. Performed in Full Range (FR) mode, the detector range is much higher than the theoretical physical range and allows to observe the actual free motion range. Performed in High Resolution (HR) mode, we observe the saturation of the detector near 30\,$\mu$m which is the HRM range lower than the free motion. The measurements in HR mode are still interesting because it is the mode commonly used for scientific measurements and give valuable information on the mass behavior when it is excentered.

So in FR mode, each mass is displaced towards one direction then the opposite. The maximum range in one direction is considered attained when the test-mass is not free of motion anymore along or about at least one of its degrees of freedom, observed as a deviation from the expected course (with respect to the stimulus) and indication of a mechanical contact between the mass and a stop.

This campaign provided insightful information on the actual mechanical configuration of the core of our instrument. Tab. \ref{table:range} summarizes all free motions deduced from the position detector. In order to be compared to the integration metrology that gives mechanical expectations, the position measurement bias due to capacitive border effect as described in Ref. \cite{liorzoucqg2} shall also be accounted for. In particular the measured value is the apparent position from a capacitive point of view which differs from actual position along $X$ axis by about 30$\mu$m for IS1-SUREF and 25$\mu$m for IS2-SUEP, deduced from ground test capacitance measurements. This leads to recenter a little IS1-SUREF and IS2-SUEP test-masses. Then, only IS1-SUEP proof-mass presents a critical free motion of only 8.1$\mu$m in the $X$ axis positive direction, which increases the noise and bias due to patch effects. However, the bias effect is quite negligible when considering a few tens of $mV$ for this phenomenon. The other axes are less sensitive to capacitive border effects. However, SUEP has a narrower cage in which the mass can move freely which is limiting for scientific measurements requiring excentered sessions (some characterization sessions or even scientific sessions could have benefited from a greater offcentering, to magnify some disturbing effects).

SUREF does meet most integration expectations though (if not the design values).

\begin{table}[h]
\begin{center}
\begin{tabular}{ m{1.5cm} c c c c c c c c }
\hline
Axis & \multicolumn{2}{c}{IS1-SUREF} & \multicolumn{2}{c}{IS2-SUREF} & \multicolumn{2}{c}{IS1-SUEP} & \multicolumn{2}{c}{IS2-SUEP} \\
 \hline\hline
 \multirow{2}{1cm}{$X ({\mu}m)$} & -78.9 & 33.7 & -36.1 & 39.2 & -32.7 & 8.1 & -55.2 & 11.3 \\
& (-34) & (39) & (-34) & (51) & (-44) & (37) & (-49) & (24) \\
 \multirow{2}{1cm}{$Y ({\mu}m)$} & -43.8 & 63.7 & -69.6 & 57.3 & -21.2 & 24.1 & -18.5 & 25.3 \\
 & (-34) & (39) & (-34) & (51) & (-44) & (37) & (-49) & (24) \\
  \multirow{2}{1cm}{$Z ({\mu}m)$} & -66.2 & 44.0 & -62.0 & 47.0 & -10.8 & 29.0 & -23.1 & 17.5 \\
 & (-34) & (39) & (-34) & (51) & (-44) & (37) & (-49) & (24) \\
 \multirow{2}{1cm}{$\Phi ({\mu}rad)$} & -2430 & 3225 & -2290 & 1120 & -6085 & 515 & -457 & 1350 \\
 & (-1932) & (2216) & (-1043) & (1564) & (-2500) & (2102) & (-1503) & (736) \\
 \multirow{2}{1cm}{$\Theta ({\mu}rad)$} & -717 & 1567 & -952 & 852 & -577 & 2028 & -519 & 1228 \\
 & (-1567) & (1797) & (-852) & (1278) & (-2028) & (1705) & (-1228) & (602) \\
 \multirow{2}{1cm}{$\Psi ({\mu}rad)$} & -1450 & 1567 & -1095 & 852 & -630 & 2028 & -617 & 1228 \\
 & (-1567) & (1797) & (-852) & (1278) & (-2028) & (1705) & (-1228) & (602) \\
 \hline
\end{tabular}
\caption{Measured and best guess from mechanical geometry (between brackets) free motion ranges in both directions per axis. The best guess free motion is computed taking into account cumulated metrological deviances without taking into account potential signs and without taking into account capacitive defects which might bias the measured position.}
\label{table:range}
\end{center}
\end{table}

On the bright side these results demonstrate that in regular operation (masses centered or with less than $5{\mu}m$ excursion during calibration), the instrument works nominally with no mechanical hindrance.

It also reveals the level of production accuracy necessary to develop and integrate future instruments capable of testing the Equivalence Principle with higher precision than what MICROSCOPE has already demonstrated.

	\subsection{Coupling between axes}

The couplings between the three translation axes and the three rotation axes of the instrument introduces a dependency between the six ideally completely independent axes. The couplings disturb the measurement by creating an additional term to compensate for the projection of the test-mass motion along the other axes. With the six axes working simultaneously, coupling from a first axis will cause the control voltage on a second axis to react, which in turn creates a coupling effect on a third axis (for a negligible effect though).

\begin{figure}[h]
        \centering
        \includegraphics[scale=0.45]{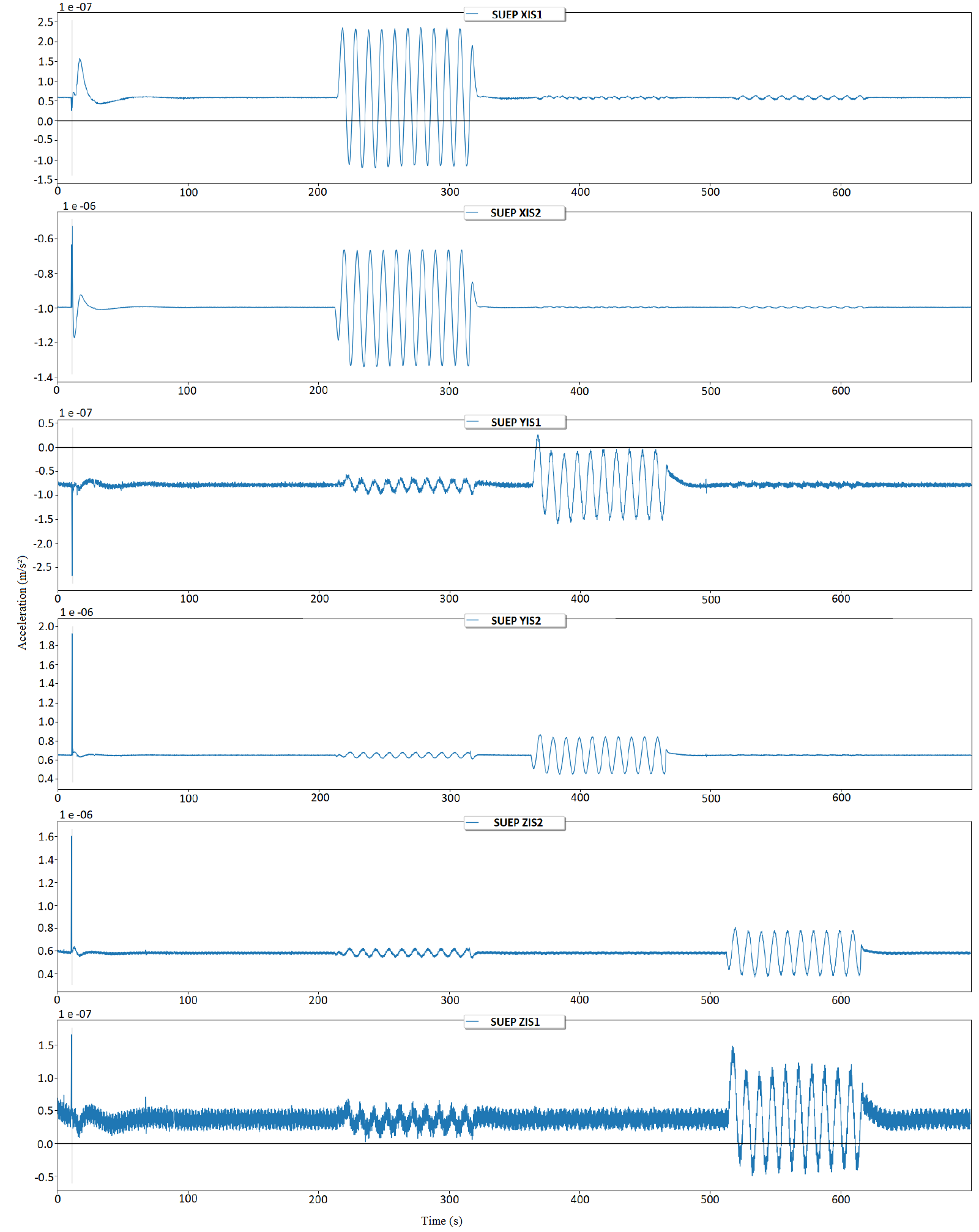}
        \caption{Accelerometer measurements during technical session 516 along the linear axes of the internal sensor (IS1) and the external sensor (IS2) for the SUEP. A sine excitation signal is introduced during $100s$ first on the $X$ axis, then on the $Y$ axis and finally on the $Z$ axis. The impact of this excitation on the other axes is due to the couplings between the axes. }
        \label{fig:couplin}
\end{figure}

Dedicated technical sessions have been performed in order to characterize the couplings in flight \cite{rodriguescqg4}. A sine signal is applied in the instrument control loop to force an oscillation of the test-mass position along the $X$ axis during $100s$ with a low frequency $f_{tech} = 0.1Hz$ outside the control bandwidth of the DFACS system which is set to compensate for the motion of the other test-mass. Then a sinusoid with the same frequency but a different amplitude is induced in the satellite motion along the $Y$ axis during $100s$, and finally along the $Z$ axis during $100s$. The impact on the measurement of the other axes (see Fig. \ref{fig:couplin}) of this excitation at a well-known frequency is studied in order to extract the coupling values. Do note that specific scientific sessions are dedicated to the estimation of linear couplings towards the $X$ axis differential acceleration though since these couplings are an important part of the calibration described in \cite{hardycqg6}.

In order to extract the coupling values, a software simulator is used that modelizes the DFACS and accelerometer servo loops. It is based on the one-axis model of the instrument control loop presented in \cite{rodriguescqg1}. This model is duplicated for the three linear axes of the internal sensor IS1 and the three linear axes of the external sensor IS2. For each axis, the DFACS control loop is represented. The DFACS control block, modelized by a fourth order transfer function and a phase delay, is set to compensate the motion of the external test-mass. The acceleration of both test-masses along each axis is injected on the other axes through the coupling parameters.

The acceleration measurement is fitted with a sinusoidal function at $f_{tech}$ with a least-square method in order to determine its amplitude along each axis for the three excitation periods. The value of the simulated coupling factors is adjusted so that the amplitude of the simulated acceleration is as close as possible to the amplitude measured in flight. The adjustment is made to determine first the impact of the acceleration along the $X$ axis on the $Y$ and $Z$ axes using the first excitation period, then $Y$ on $X$ and $Z$ using the second period and finally $Z$ on $X$ and $Y$ using the third one. As each axis impacts the others, the adjustement process is repeated until convergence of the coupling factor values. The results are presented in Tab. \ref{table:linlincoup}.

\begin{table}[h]
\begin{center}
\begin{tabular}{ m{3cm} c c }
\hline
Linear to linear coupling factor & IS1-SUEP & IS2-SUEP\\
 \hline\hline
 $X \to Y$ & $3.2\times10^{-2}\pm1\times10^{-3}$ & $-0.16\pm5\times10^{-3}$ \\
$X \to Z$ & $3.4\times10^{-2}\pm2\times10^{-3}$ & $-0.17\pm1\times10^{-2}$ \\
$Y \to X$ & $-1.3\times10^{-2}\pm2\times10^{-3}$ & $-1.3\times10^{-2}\pm2\times10^{-3}$ \\
$Y \to Z$ & $2.2\times10^{-2}\pm5\times10^{-3}$ & $1.8\times10^{-2}\pm1.9\times10^{-2}$ \\
$Z \to X$ & $-2.9\times10^{-2}\pm1\times10^{-3}$ & $-3.1\times10^{-2}\pm2\times10^{-3}$ \\
$Z \to Y$ & $-2.5\times10^{-2}\pm2\times10^{-3}$ & $-1.9\times10^{-2}\pm8\times10^{-3}$ \\
 \hline
\end{tabular}
\caption{Linear to linear coupling factors determined by an iteration process for the internal sensor (IS1) and the external sensor (IS2) for the SUEP and associated least-square error. \hl{Additionally to the statistical errors indicated in the table}, a few \% error of the simulation models also has to be taken into account and leads to a systematic error of the order of 0.05 to be added to the least-square error.}
\label{table:linlincoup}
\end{center}
\end{table}

This characterization is useful to use the acceleration measurement in common mode for other applications of the data. In differential mode, the calibration sessions are much more accurate and show in particular a very small coupling from radial axes toward the $X$ axis (less than 10$^{-4}$) \cite{hardycqg6}. The coupling factors in Tab. \ref{table:linlincoup} are established at 0.1\,Hz. The consequence is that these estimated couplings have to be corrected from the transfer function to get estimated values near EP frequencies of interest [0.0009\,Hz - 0.0065\,Hz]. Furthermore, the DFACS loop has been simplified and its parameters estimated on the basis of a fit response to particular signals. The error of the model should represent a maximum of 0.05 (\hl{this is an absolute error on the couplings which are unitless, corresponding to the dispersion of the coupling values from $Y$ or $Z$ axes toward another when established at frequencies of stimuli other than 0.1\,Hz and model error in the simulator, especially at frequencies at drag-free control system bandwidth end}). When using the calibration sessions, the couplings in differential mode are compatible with Tab. \ref{table:linlincoup} results within the presented accuracy. The strong coupling of acceleration along $X$ upon $Y$ or $Z$ is explained by the impact of the applied voltage for $X$ electrodes to compensate the bias on the stiffness of $Y$ or $Z$ control forces. The equations are detailed in Ref. \cite{liorzoucqg2} and can be summarized for the electrostatic control acceleration along $Y$ as an example :
\begin {equation}
\Gamma_y \sim k_1 v_y + \alpha_y (V_{py}^2+v_y^2+v_\psi^2) y +\alpha_z (V_{pz}^2+v_z^2+v_\theta^2) y+\alpha_x (V_{px}^2+v_x^2) y+\alpha_\phi (V_{p\phi}^2+v_\phi^2) y
\label {forcey}
\end{equation}
where $k_1$ is the physical gain for the $Y$ control, $V_{pw}$ is the difference of reference voltages between test-mass and electrodes controlling the $w$ degree of freedom, $\alpha_w$ is a geometrical factor, $v_w$ is the voltage calculated by the digital controller for each degree of freedom and $y$ the displacement along $Y$ as the one resulting from the bias of the electronics. In the case of SUEP, the gold wire stiffness is quite large and leads to have a control voltage $v_x = v_{x_0}+v_{x_t}$ while on others axes $v_w$ has a mean value close to zero. At first order, the acceleration along $Y$ presents a term $2\alpha_x  v_{x_0} v_{x_t}y$. That means that an applied acceleration along $X$ will induce a controlled voltage variation of $v_{x_t}$ proportional to the acceleration which will be also measured along $Y$ with the scale factor $2\alpha_x  v_{x_0}y$. As a small offcentering exists along all axes, this effect is not so negligible as shown in Tab. \ref{table:linlincoup}. \hl{In spite of the very accurate manufacture of the different mechanical parts, the integration process drives the budgets of misaligments and miscenterings which explain the tendencies shown in the table for all couplings of $X$ toward all others axes. Even with small misalignment values (a few 10$^{-3}$\,rad with respect to the main axis) and miscentrings (5$\mu$m), couplings, potentially important, are created by cross electrostatic stiffnesses coupled to strong acceleration DC biases along $X$ (due to the gold wire).}

	\subsection{Angular to linear axes coupling}

The measurement of the couplings from the angular axes to the linear axes has been performed on each SU with a different approach. The principle is to induce an angular oscillation in the instrument environment and observe its influence on the linear acceleration measurements. The amplitude of the sine signal along the linear axis is estimated by a least-square method and compared to the estimated oscillation signal to finally give the angular to linear axis coupling for that axis (the results are presented in Tab. \ref{table:anglincoup}).

With the drag-free and attitude control centered on the external mass, the satellite is summoned to perform a sine oscillation motion around each of the three axes consecutively. During each oscillation, we estimate the amplitude of that motion through the external mass acceleration around the oscillation axis. This acceleration is then used as a base of comparison with each linear axis acceleration measurement of the internal mass, which actually is the differential acceleration along that axis given the drag-free setting. This measurement $\Gamma_{meas}$ is corrected from the effect of the relative mass excenterings $\Delta$ coupled to the satellite angular motion and the Earth gravity gradient tensor. Once this term $[(T-In)\Delta]$ is corrected from, the actual coupling $C_{j,i}$ for each axis is evaluated (see Eq. (\ref{eq:anglin})).

\begin{equation} \label{eq:anglin}
\ \Gamma_{meas,i}=\Gamma_{0,i}+C_{j,i}\dot{\Omega}_{osc,j}+[(T-In)\Delta]_{i}\
\end{equation}
where $i$ designates a linear axis, $j$ an angular axis and $\dot{\omega}_{osc,j}$ the angular acceleration due to the satellite motion around axis $j$.

\begin{table}[h]
\begin{center}
\begin{tabular}{ c c c }
\hline
$(m/s^2)/(rad/s^2)$ & SUREF & SUEP \\
 \hline\hline
$\Phi \to X$ & $-2.4\times10^{-5}\pm1.6\times10^{-6}$ & $9.6\times10^{-6}\pm9.2\times10^{-7}$ \\
$\Theta \to X$ & $9\times10^{-7}\pm1.7\times10^{-6}$ & $-1.6\times10^{-5}\pm1.0\times10^{-6}$ \\
$\Psi \to X$ & $6\times10^{-7}\pm1.7\times10^{-6}$ & $-2\times10^{-7}\pm1.0\times10^{-6}$ \\
$\Phi \to Y$ & $4.53\times10^{-4}\pm2.4\times10^{-6}$ & $-2.44\times10^{-4}\pm1.7\times10^{-6}$ \\
$\Theta \to Y$ & $-5.1\times10^{-6}\pm1.7\times10^{-6}$ & $-1.9\times10^{-5}\pm2.1\times10^{-6}$ \\
$\Psi \to Y$ & $-1.05\times10^{-4}\pm1.6\times10^{-6}$ & $-8.92\times10^{-4}\pm1.9\times10^{-6}$ \\
$\Phi \to Z$ & $7.60\times10^{-4}\pm1.5\times10^{-6}$ & $1.50\times10^{-3}\pm3.3\times10^{-6}$ \\
$\Theta \to Z$ & $-1.7\times10^{-5}\pm1.5\times10^{-6}$ & $1.09\times10^{-3}\pm5.0\times10^{-6}$ \\
$\Psi \to Z$ & $5.5\times10^{-5}\pm1.6\times10^{-6}$ & $7.0\times10^{-5}\pm3.1\times10^{-6}$ \\
 \hline
\end{tabular}
\caption{Angular to linear differential couplings computed by comparison of least-square method estimations of acceleration measurement during dedicated satellite oscillation sessions.}
\label{table:anglincoup}
\end{center}
\end{table}

\hl{In the analysis of the measurements directly dedicated to the estimation of the E\"otv\"os parameter, the coupling term can be estimated using the measured angular acceleration and the impact of the measurement error of this angular acceleration can be computed} \cite{rodriguescqg1, hardycqg6}.


\section{Conclusion}

While not a direct part of the scientific extraction of the EP signal, the characterization of the instrument helps better understanding the instrument and its measurements \hl{and estimating the systematic effects as done in} \cite{hardycqg6} \hl{following the budget process explained in} \cite{rodriguescqg1}. \hl{It also ensures that, for the data exploitation, the instrument and its servo-control are functioning correctly, the control laws optimized.} The consequences of this knowledge acquired thanks to the characterization range from orienting the scenario or modifying the instrument parameters (during the mission) to improve its behavior to simply going towards the validation of the EP test.

\hl{In between, the stiffness estimation may contribute to the total noise, especially with respect to the systematics, at the order of a fraction of 10$^{-16}$m/s$^2$. The noise evolution conveys the behavior, quite stable, of the instrument over the mission course; this information is taken into account in the data exploitation. The quadratic factor seems to vary a lot during the mission when observed through the estimation of the differential quadratic factor} \cite{hardycqg6}. \hl{The quadratic factor estimated through the method presented in this paper, which is only one measurement point, is therefore difficult to apply over all the mission}. \hl{In the data exploitation, a capping value is rather used. The free motion ranges are not used directly. The couplings however, do represent one error item in the systematics} \cite{hardycqg6}. \hl{In the end, we check various characteristics to verify they are as expected; some of these, such as stiffnesses and couplings, are used in the systematics estimation. Beyond the validation of the instrument operation and the EP performance evaluation, this characterization also serves to give some key numbers for future uses of the MICROSCOPE data for any other application.}


\ack
The authors express their gratitude to all the different services involved in the mission partners and in particular CNES, the French space agency in charge of the satellite. This work is based on observations made with the T-SAGE instrument, installed on the CNES-ESA-ONERA-CNRS-OCA-DLR-ZARM MICROSCOPE mission. ONERA authors’ work is financially supported by CNES and ONERA fundings.
Authors from OCA, Observatoire de la C\^ote d’Azur, have been supported by OCA, CNRS, the French National Center for Scientific Research, and CNES. ZARM authors’ work is supported by the DLR, German Space Agency,  with funds of the BMWi (FKZ 50 OY 1305) and by the Deutsche Forschungsgemeinschaft DFG (LA 905/12-1). The authors would like to thank the Physikalisch-Technische Bundesanstalt institute in Braunschweig, Germany, for their contribution to the development of the test-masses with funds of CNES and DLR.

\section*{References}
\bibliographystyle{iopart-num}
\bibliography{biblimscope}

\appendix
\section{Controller performance}
	
The controller performance has been evaluated on ground before the mission using a software simulator of the instrument dynamics (see Fig. \ref{fig:looplink}) on the basis of the schema in Fig. \ref{fig:PID}. The instrument model, designed in Matlab\textregistered{}/Simulink, simulates the motion of a test-mass along each axis, and the rotation about each axis. \hl{The simulator handles all 6 degrees of freedom, although only one is presented here for clarity, with couplings linking one to another. In particular, couplings measured in flight are reintroduced afterwards in the simulator.} The input of the simulator is the external non-gravitational acceleration applied on the test-mass. A double integrator computes the test-mass displacement or rotation along the chosen axis, taking into account the potential rebounds of the mass if it hits the stops. The simulation of a velocity step as specified for a piece of debris is performed by adding a signal in the double integrator box, between the two integrators.

\begin{figure}[h]
        \centering
        \includegraphics[scale=0.6]{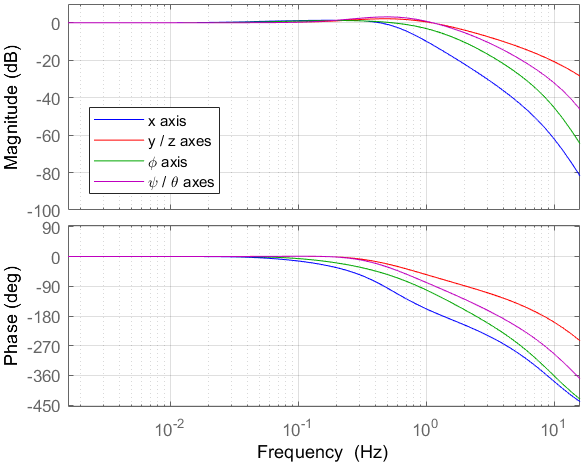}
        \caption{Bode diagram of the closed-loop transfer functions of the accelerometer along the different axes. The transfer functions are identical for the four test-masses}
        \label{fig:bode}
\end{figure}

The capacitive detector is modeled by a first order transfer function followed by a saturation representative of the electronics and provides a voltage proportional to the measurement of the test-mass position and converted to a digital number by an analog-to-digital converter \cite{liorzoucqg2}. The PID block uses this measurement to compute the voltage to be applied onto the electrodes in order to control the test-mass motionless. Its output is converted into an analog voltage amplified by the Drive Voltage Amplifier (DVA) represented by a first order transfer function and a saturation. The computed voltage is applied to the electrodes surrounding the test-mass and results in an additional electrostatic force. The stiffness of the instrument (see section \ref{section:stiffness}) is modeled by a gain introducing an effect on the test-mass position offset in the electrostatic actuation. The closed loop transfer function of the instrument, computed by the software simulator, is represented in Fig. \ref{fig:bode}. Its corresponding characteristics are presented in Tab. \ref{table:bode}. The transfer function at low frequency has been optimized in order to keep a phase shift lower than 28$^\circ$ at 0.1Hz for the DFACS stability needs. The noise and bias of the instrument use less than 50\% of the measurement and control bandwidth (or 54\% for the $\Psi$ and $\Theta$ axis of the internal masses), therefore suppressing any risk of autosaturation of the instrument.

\begin{figure}[h]
        \centering
        \includegraphics[scale=0.25]{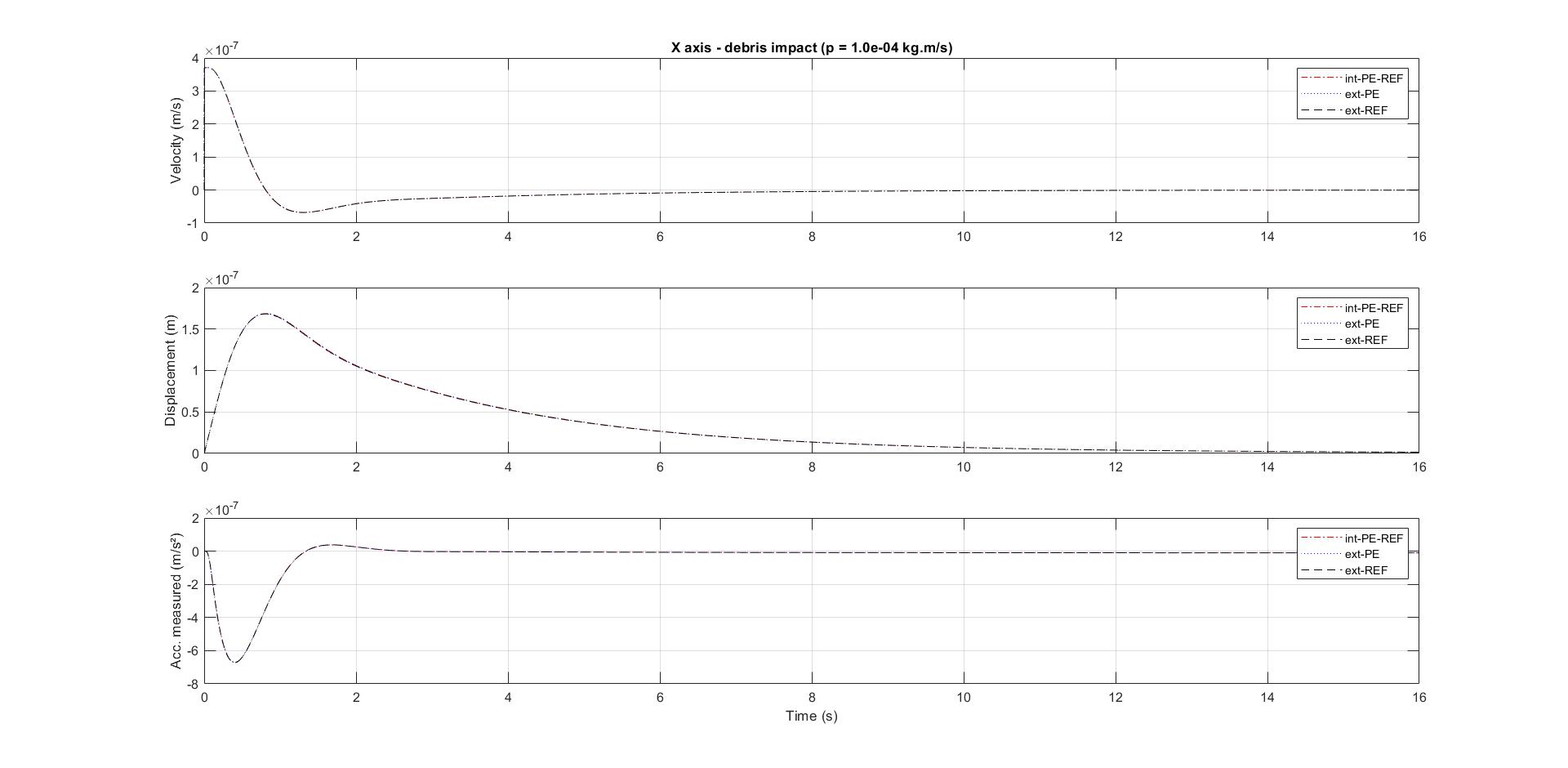}
        \caption{Step response of the accelerometer $X$ axis control loop when undergoing an impact with a $10^{-4}Ns$ momentum}
        \label{fig:stepanswer}
\end{figure}

Finally, Fig. \ref{fig:stepanswer} shows the response of the instrument reacting to a step input of $3.7 {\times}10^{-7}m/s$ which corresponds to the impact of a micrometeorite with a momentum of $10^{-4}$N\,s considering the mass of the satellite: such event was expected to occur twice a month by CNES and thus during most of the science sessions. The response to the maximum momentum of $1.4\times{}10^{-3}$N\,s was also simulated but corresponds to a smaller probability (twice a year). As specified the test-mass does not reach the stops. SUEP features the same performance as SUREF except for the autosaturation of IS2-SUEP being half the IS2-SUREF values because of the bias voltages applied on electrodes \cite{liorzoucqg2} that modify the range.

\begin{table}[h]
\begin{center}
\begin{tabular}{@{}|l|cccc|}
\hline
& \multicolumn{4}{c|}{SUEP and SUREF} \\

\multicolumn{1}{|r|}{Axis} & $X$ & $Y/Z$ & $\Phi$ & $\Theta/\Psi$  \\
 \hline\hline
Bandwidth (Hz) & 0.62 & 1.8 & 1.0 & 1.6  \\
Overshoot$^{(a)}$ (dB)  & 1.3 & 1.0 & 1.2 & 1.4  \\
Gain margin (dB) & 16.4 & 19.2 & 15.0 & 14.9  \\
Phase margin ($^{\circ}$) & 53.9 & 60.6 & 60.3 & 47.9 \\
Phase shift ($^{\circ}$)$^{(b)}$ & 27.3 & 13.1 & 16.4 & 12.9   \\
\hline
Autosaturation$^{(c)}$  &  &  &  &  \\
IS1  & 9.3 & 47.0 & 44.2 & 53.9 \\
IS2-SUEP / SUREF & 5.0 / 6.3 & 12.5 / 27.9 & 4.4 / 9.5 & 11.7 / 20.4  \\
(\% of the range) & & & & \\
 \hline
\end{tabular}
\caption{Performance of the instrument control loop computed by the simulation software: bandwidth is calculated at -3dB; (a): the maximum overshoot gain in the frequency range [0.005Hz-0.2Hz];  (b): the phase shift is considered at 0.1Hz of the accelerometer output for the DFACS, it takes into account the PID transfer function, the anti-aliasing filter and the mean filter; (c): the saturation in the control loop is evaluated as the sum of the bias and of the 5${\sigma}$-noise. IS2-SUEP has better phase shift and autosaturation performance than its SUREF counterpart.}
\label{table:bode}
\end{center}
\end{table}

\end{document}